\documentclass[12pt]{article}
\usepackage{amsmath}
\usepackage{latexsym}
\usepackage{bbm}
\usepackage{color}
\usepackage{graphicx}
\def\ket#1{\left| {#1}\right\rangle}
\def\half{\frac 1 2}
\def\re{\hbox{Re\ }}

\newdimen\mpindentsz\mpindentsz=\the\parindent 
\def\mpindent{\leavevmode\hbox to \mpindentsz{\phantom{a}\hss}}

\begin{document}
\begin{center}
{\Large
Reminiscence on the Birth of String Theory\footnote{\normalsize
  Invited Contribution to  
``The Birth of String Theory" Commemorative Volume}\\[12pt]
Joel A. Shapiro\\[12pt]}
{\large \em Department of Physics and Astronomy\\
Rutgers University, Piscataway, NJ, 08854-8019}
\end{center}
\vskip 24pt
\begin{center}
  Abstract
\end{center}

These are my personal impressions of the environment in which string
theory was born, 
and what the important developments affecting my work were during the
hadronic string era, 1968-1974. I discuss my motivations and concerns at the
time, particularly in my work on loop amplitudes and on closed strings.

\vskip 1in

\section{Introduction}

It is not unusual in theoretical physics for conceptual frameworks to 
ride roller-coasters, but few have had as extreme highs and lows as in the
history of string theory from its beginnings in 1968 to the present. In fact,
string theory was so dead in the mid to late '70's that 
it is a common assumption of many articles in the popular press, and
of many younger string theorists, that the field
originated in the '80's, completely ignoring the period we are celebrating
here, which is primarily 1968-74. 

So it was pleasantly surprising to be invited to reminisce about the early
days of string theory. Research results from that era have been extensively
presented and reviewed, so I will try to give my impression of the atmosphere
at the time, and what questions we were trying to settle, rather than review
the actual results.

\section{The Placenta}

In the mid '60's, the framework for understanding fundamental physics was
very different from what it is now. We still talk about the four fundamental
interactions, but we know that the weak and electromagnetic interactions
are part of a unified gauge field theory, that strong interactions are 
also described by a gauge field theory which might quite possibly unify with
the others at higher energy, and that even general relativity is a form
of gauge field theory.  In the 1960's things were very different. Not only
were the four interactions considered to be of completely different natures,
but for the most part the physicists who worked on them were divided
into groups by the  
interactions on which they worked.  Of course, every budding particle
theorist learned QFT and how wonderfully successful it was in treating
QED. But one also learned how these perturbative methods could not be
used for strong interactions because the coupling constant was too
large, and that for the weak interactions one could only work at the Born
approximation, because all existing field theories for the weak
interactions were non-renormalizable. 
So particle theorists were divided into separate groups: one working
on strong interactions, one on weak interaction phenomenology, 
and one doing high order, esoteric QED calculations. 
Each group had very different techniques and styles.

Even more removed from the world of a strong interaction physicist was
the fourth interaction, gravity, which was studied, if at all, by 
general relativists\footnote{How separated general relativity and particle
physics were in the '60's is discussed by David Kaiser \cite{Kaiser}. He
argues that funding cuts in particle theory in the late '60's and '70's 
played a large role in the subsequent bringing together of particle theorists
and general relativists.}. When, as a very na\"ive graduate student who knew 
nothing of the fields of physics research (I had just received my ScB in 
Applied Mathematics), I was asked by my future advisor what I might be 
interested in, I replied ``unified field theory''. Nonetheless, it was
never suggested that I take a course in general relativity! 

So the context into which string theory was born was not so much theoretical 
fundamental physics or even particle theory, but rather strong interaction
theory/phenomenology. The principal recent successes in that field had
been in searching for patterns and fitting simple models\footnote{
I am using ``model'' and ``theory'' with a distinction that is perhaps not
generally accepted. To me, a theory is a comprehensive approach to explaining
part of physics in a way which will at least have features which are 
fundamentally correct,
while a model tries, with less ambition, to fit aspects of the data, but 
cannot be taken as the fundamental truth, 
even as an approximation of the truth. Thus
QED, QCD, and general relativity are theories, even though the last
clearly needs modification to include quantum mechanics, while the 
interference model, DHS duality\cite{DHS}, and my thesis are models. The Dual 
Resonance Model might be taken to have evolved into a theory when we 
started calculating unitary corrections in the form of loop graphs.}
to scattering data.
Scattering cross-sections were dominated by resonance peaks and the high
energy asymptotic behavior described by Regge trajectories. A huge number
of particles and resonances had been found and were listed in the
particle data tables \cite{pdg67}. The organization of these
particles into (flavor) SU(3) multiplets was the most impressive thing
understood about the strong interactions. 

The quasi-stable hadrons and the resonances fell beautifully into 
patterns which could be understood by treating baryons {\em as if}
they consisted of three quarks and mesons of a quark and an antiquark.
Even though this very successfully described the dominant experimental
observations, theorists were very reluctant to think of the quarks as
real constituents of hadrons.

Fits to the data were done by treating the scattering amplitude as a
sum of resonance production and decay, together with an additional
contribution due to the exchange of the same particles in the form of
Regge poles to describe the high energy behavior. This sum was called
the interference model.
But the experimentalists kept
finding more and more resonances, and they were joined by phase-shift 
analysts. It soon appeared the 
sequence of resonances continued indefinitely to higher masses and
spins, in 
what clearly looked like linearly rising Regge trajectories. In fact, 
my Ph.~D.~thesis\cite{JSthesis} was a very na\"ive non-relativistic
model using PCAC, 
which rather successfully explained the experimental\cite{pdg67} $\pi$-nucleon
decay widths of a tower of five $\Delta$ resonances with spins ranging
from 3/2 to 19/2. Unfortunately the top two of these resonances have 
subsequently dissolved\cite{pdg06}.
This infinite sequence of resonances suggested the idea of duality
\cite{DHS}, that the amplitude  
could be described {\em either} in terms of a sum of resonances {\em
  or} in terms of a series of Regge poles. The possibility that a 
scattering amplitude $A(s,t)$ could be given as an sum of resonant 
poles in $s$ or alternatively as a sum of Regge poles in $t$ caused great 
excitement, but also skepticism that such a function could exist. 

\section{Conception and the Embryonic Period}

Thus it seemed miraculous when Veneziano \cite{Venez} discovered that
Euler had given us just such a function in 1772, to describe the 
$\pi\pi\rightarrow \pi \omega$ scattering amplitude. This paper
arrived at the Lawrence Radiation Lab in Berkeley in the summer of
1968 while I was away on 
a short vacation, and I returned to find the place in a whirlwind of 
interest. Everyone had stopped what they were doing, and were asking 
if this idea could be extended to a more accessible interaction, such as
$\pi\pi\rightarrow \pi \pi$. I suggested the very minor modification
necessary to remove the tachyon, 
$$ \frac{\Gamma(-\alpha(s))\Gamma(-\alpha(t))}{\Gamma(-\alpha(s)-\alpha(t))}
\longrightarrow \frac{\Gamma(1-\alpha(s))\Gamma(1-\alpha(t))}
 {\Gamma(1-\alpha(s)-\alpha(t))},$$
and Joel Yellin and I investigated 
whether this could be taken as a realistic description \cite{pipisy}
for $\pi\pi$ scattering. It had a lot of good qualitative features,
including resonance dominance, regge behavior, and full duality.
We were forced to have exchange degeneracy between the $I=0$ and $I=1$
trajectories, which was well fit by the data. 
We noticed the problem that such an amplitude can wind up with ghosts,
with a negative decay width for the $\epsilon'$, the $0^+$ partner of
the $f$, but also that this problem disappeared if the $\rho$
trajectory intercept exceeded 0.496, very close to the value of 0.48
which we got from fitting the low energy phase shifts. 
A much more serious problem was that we predicted a $\rho'$ degenerate
with the $f$, which seemed to be ruled out by experimental data. That
the simplest function did not produce a totally acceptable model was
discouraging, especially to Yellin, although we realized that there was
no compelling reason not to add subsidiary terms to the simple ratio
of gamma functions, except that to do so removed all predictive power!
This convinced Yellin that he didn't want to coauthor the fuller 
version\cite{pipipr} of our paper. But Lovelace\cite{Lovelacepp}, who 
independently discovered the same amplitude, managed to do a favorable 
comparison to experiment.

There were a number of papers attempting to do phenomenology with
dual models, mostly describing two-body scattering processes.
In general the results had, as did our paper, nice
qualitative features but unsatisfactory fitting of the data.  At the 
same time, the formal model was becoming much more serious, as great progress 
was made in extending the narrow resonance approximation amplitude,
first to the 5 point function
\cite{BardRueggA} and then the $n$-particle \cite{BardRueggB,CHM47,CHMT,
GoebelSak}
amplitudes.
A very elegant formulation of these amplitudes was given
by Koba and Nielsen\cite{KobaN,HBNielsen,KobaNb}, in which the external
particles correspond
to charges given by their momentum, entering on the boundary of a unit circle,
and the amplitude is given by an integral, over relative positions of the
particles, of the two-dimensional electrostatic energy. 
Here the conformal invariance was seen to play a crucial role, and in 
particular the M\"obius invariance explained the cyclic symmetry. 
From the $n$-point amplitude for ground state particles
one could factor in multiparticle channels to extract the scattering
amplitudes for all the particles which occurred in intermediate states,
determining all amplitudes in what could be considered 
the equivalent of the tree approximation in a Lagrangian field theory. 
We took the attitude that the particles of 
the theory should be all and only those which arose from $n$-point
scattering amplitudes of the ground state particles, as intermediate
states in the n-point tree function. The amplitude for an arbitrary 
particle $X$ connected to $p$ ground states
could be found by 
factoring the $p+q$ ground state amplitude \cite{BardMandel}, and amplitudes
involving more arbitrary states could come from factoring that. Thus one
could determine, in the tree approximation, the arbitrary $n$-particle 
amplitude. In a sense, this was a form of bootstrap, as the set of particles
generated as intermediate states were added to form a consistent set, with
the same particles as intermediate states as were considered external states.

\section{Birth of String Theory}

Of course a set of tree amplitudes is not a unitary theory. In perturbative
field theory, the Feynman rules are guaranteed to implement unitarity
by specifying loop graphs whose discontinuities give the required
sum over intermediate states, because these all come from a Hermitian 
lagrangian. The possibility of 
advancing dual models to a unitary theory became possible once 
we had the tree amplitudes for arbitrary single particle states,
as one could sew together the loop 
graphs to give a perturbative (in the number of loops) theory
satisfying unitarity.
In perturbative quantum field theory, loop
graphs give the appropriate contributions to the optical theorem, satisfying
the unitarity of the $S$-matrix. Bardak\c c\i, Halpern and I (BHS)\cite{BHS} 
defined the 
one loop graph by the requirement of two-particle unitarity. An earlier
attempt (KSV)\cite{KSV} defined the planar loop graph by extending duality
to the internal legs, which gave most of the factors in the
loop integrand.  But to get the full expression, the one-loop amplitude for 
$n$ ground-state particles $\sigma$ should be required to have the correct 
two particle discontinuity, a sum over all possible two-particle intermediate
states. Starting with the tree amplitude for $n$ $\sigma$'s plus $X(p)$
plus $X(-p)$, and summing over 
\phantom{mmmmmmmmmmmmmmmmmmmm}
\medskip\vskip -11pt\noindent
\begin{minipage}[h]{4.6in}

all possible states $X$ and momentum $p$, 
as shown by the stitch marks here, one is summing not only over $X$ but 
also over all particles in the left arm, because those are all included in
the tree graph. 

\mpindent Of course we called this process ``sewing'', which led to an amusing
battle with Sy Pasternack, the editor of Physical Review, on a 
subsequent paper\cite{HKS}. Pasternack thought he needed to uphold a
certain formality, 
and was responsible for ``pomeronchukon'' rather than ``pomeron''. He wrote
us a very witty letter\cite{PastKnit} arguing that ``sewing'' would
lead inevitably to  
``weaving'', ``braiding'', ``darning'', ``knitting'' and ``sew-on''. 
We
\phantom{mmmmmmmmmmmm}
\end{minipage}
\quad
\begin{minipage}[h]{1.7in}
\includegraphics[width=1.7truein]{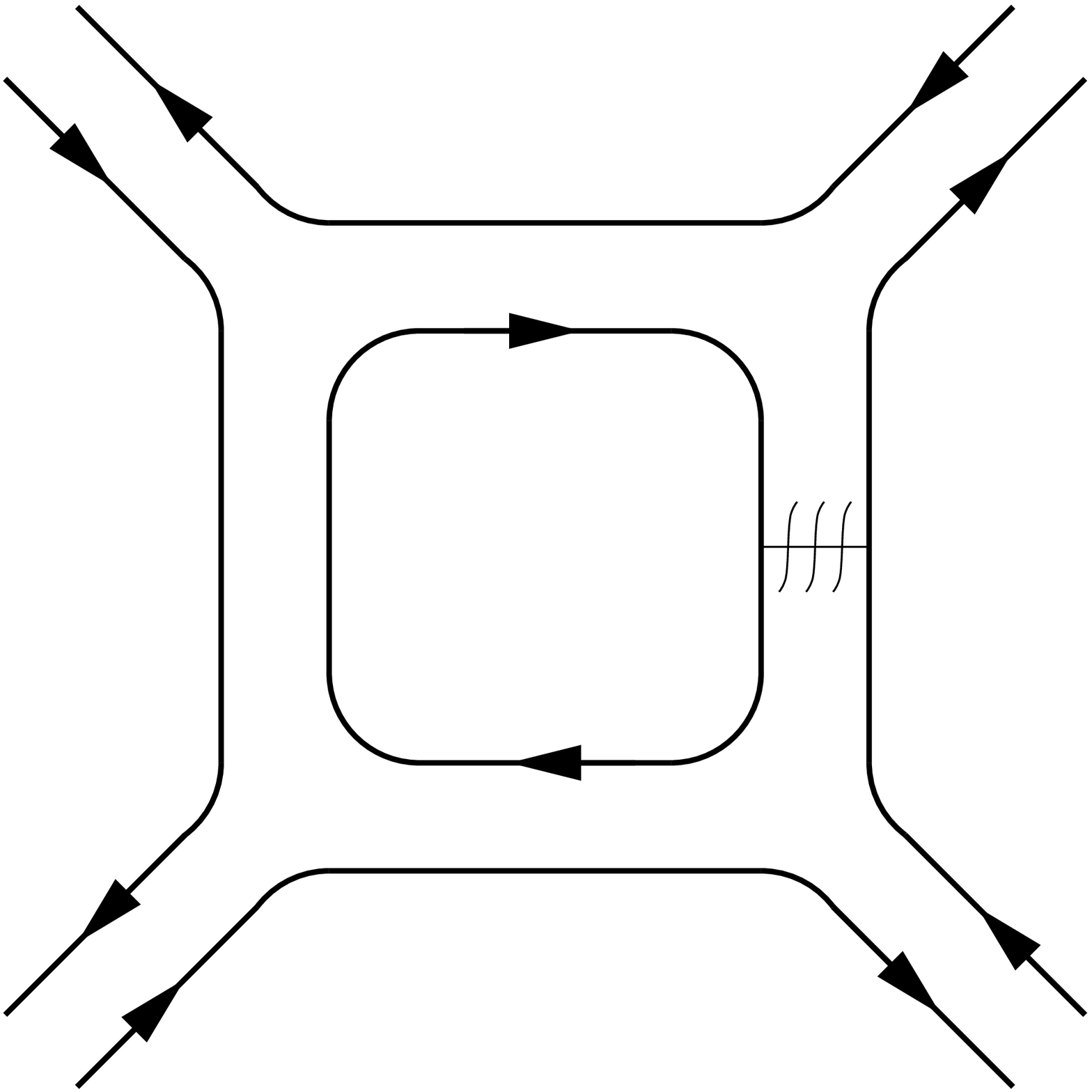}
\end{minipage}

\vskip-10pt\noindent
objected, however, that the actual thread lines were shown in the
figures.
Redrawing figures in those days was a major undertaking.
That won the argument.

I should point out that at the time, our description of the
intermediate states and the amplitudes was quite clumsy, 
using the rather messy techniques
of the Bardak\c c\i-Mandelstam factorization \cite{BardMandel}.
While we were working on deriving the loop graph,
Nambu \cite{Nambu},  and
Fubini, Gordon and Veneziano \cite{FGV}
were developing the elegant operator formalism, in
which the states of the system are 
described by harmonic oscillator excitation operators  $a_n^{\mu 
\,\dagger}$,  $n=1...\infty$ acting on a ground state $\ket 0$. Here
$n$ corresponds to nodes on a string and we have a Lorentz index $\mu$.
The amplitudes can then be written as a matrix element with vertex functions
for each external particle, and propagators integrated $\int_0^1 du$
over an internal variable $u\sim e^{-\tau}$, where $\tau$ 
acts like a time describing how long an intermediate state propagates.
Thus resonance poles in a tree, or two particle intermediate states
in a loop, come from the $\tau\rightarrow\infty$, $u\rightarrow 0$
limit for the corresponding propagator.   This formalism
made the calculations much easier. It enabled
the authors of KSV to discover independently from
us the extra factors that get two particle unitarity correct, except for
spurious states.
The new formalism was so superior\cite{ABG} that few people were encouraged to
read our paper, and I am still grumpy about that.

The operator formalism made clearer two problems that had already been 
vaguely seen.
In this formalism, the amplitudes appear to lose the M\"obius 
invariance, but the amplitudes do not, due to the existence of Ward
identities. That is, there are spurious states, 
combinations of excitations which decouple from all $n$ ground state 
amplitudes,
and therefore by our philosophy should not be included among the states.
Secondly, the time-like creation operators create ghosts, particles
with negative widths, which clearly should not be there. The set of these 
ghosts produced by the lowest node operator, $a^{0\,\dagger}_1$, were
precisely those that could be exorcised by those Ward identities
known at the time. Much of our effort in BHS involved excluding these
spurious states from the loop. Of course the higher time-like modes 
$a^{0\,\dagger}_n$, $n>1$ also produce ghost states, but these
were also exorcised by the Ward identities found later
by Virasoro\cite{ViraLn}.

One unpleasant feature of the planar loop amplitude we constructed
from two-particle unitarity was the presence of additional factors
$\prod_r (1-w^r)^{-D}$, where $w=\prod u_j$ is the product of the $u$
factors of all the internal propagators, and $D$ is the dimension of
space-time, which at that point we simply wrote as 4. The
discontinuities we were building into the loops come from several
intermediate state $u_i$'s going to zero, so only the $w\rightarrow 0$
endpoint contributes, but the natural integration
range for $w$ was from 0 to 1.
Of course at $w=1$ this factor has
extremely bad behavior. 
Eliminating the one set of spurious states known about at the time
eliminated just one power of $(1-w)^{-1}$, which didn't help much. 
Later, Virasoro \cite{ViraLn} discovered that if the Reggeon intercept
$\alpha(0)=1$,  
there was an infinite set of generators of spurious states, and
eliminating those gets rid of all the ghosts (for $D\leq 26$), and one
full set of $\prod_r (1-w^r)^{-1}$. Still, there is a very serious
divergence as $w\rightarrow 1$, which will turn out to be connected to
the Pomeron/closed string.
Before that was realized, there was speculation about whether this 
divergence was removable, and whether the two particle
discontinuity had the expected two-Reggeon cut asymptotic
form\cite{2partDisc} as $s\rightarrow \infty$. 

It should be mentioned that this effort to raise dual models to the 
same level of legitimacy as perturbative QFT was a departure which
made many uncomfortable. Strong interaction theorists had been divided
into field 
theorists and S-matrix folk, and dual models were generally considered
the domain of S-matrix types, but here they were adopting the moral
values of a field theorist, even if the context was different. The 
phenomenologically inclined thought it would be better to simply assign
imaginary parts to the regge trajectories in the dual amplitudes to go
beyond a narrow resonance fit. As there had been no real data-fitting
successes, many had great skepticism about the value of dual model
research. One such skeptic asked me why I would work on something so
unlikely to be the real physical truth. I recall saying that even 
though the probability that dual models would be part of the real
answer was small, perhaps 10\%, at least there was a chance of working
towards the truth, while fitting elastic scattering data to Regge
poles, to me, seemed not to have any chance of leading to fundamental
physical understanding.

I mention this because in 1987, in Aspen where string theory was 
the superhot theory of everything, I asked some of the younger
researchers what their estimate was, of the probability that string 
theory would be part of the the true theory of physics, and was rather
astounded to hear answers upwards of 50\%. 

Anyway, let's get back to the construction of loop graphs for a
complete, unitary dual resonance theory. This was a very active field.
Neveu and Scherk \cite{NeveuScherk} used their superior French
mathematical education to express the planar loop in terms of elegant
Jacobi $\theta$ functions, enabling them to extract the divergent 
behavior. The operator formalism  \cite{FV, FGV, SusskindHS} made tractable
the calculation of nonplanar loops \cite{KKSV,GNSS,KakuThorn} and
multiloop amplitudes\cite{KakuYua}. Abelian integrals were used by
Lovelace\cite{Lovelace70,Aless71,Lovelacedrc}, who suggested that
experimentalists deprived of funding for higher energy machines could
``still construct duality diagrams in tinfoil and measure induced charges''
as their contribution to understanding particle physics.

\section*{Closed Strings}

My second post-doc appointment was at the University of Maryland, which had a 
pleasant and active high-energy theory group, but no one doing dual
models. I felt quite isolated, and while I was able to write some
technical papers\cite{genFact,JaSnonorient}, I missed the stimulating
environment I had had in Berkeley. In particular, the first of these papers
was a rather misguided attempt to get rid of the spurious states given 
by the Ward identity with $L_1$, before I became aware that Virasoro
had found, for the ``unrealistic choice of $\alpha(0)=1$'', that there 
was an infinite set of such Ward identities, enough to get rid of all
the ghosts produced by $a^{0\,\dagger}_n$. 
Fortunately I was free to visit
Berkeley and Aspen during the summers. During my Berkeley stay in 
1970 I spent a day at SLAC, where in a discussion with Nussinov and
Schwimmer, they asked me a very interesting question. At the time,
the n-point Veneziano formula was best described by the Koba-Nielsen
picture of external charges (or currents) on the circumference of a
disk. The integrand could be interpreted as the electrostatic energy
of the charges or as the heat generated by the currents, inside the 
disk. There was also much interest in this being an approximation of
very complex Feynman diagrams called fishnets within the disk. The 
question Nussinov asked is  what would happen if the external particles,
instead of residing on the circumference of a disk, where on the surface of
a sphere. Nussinov answered his own question with ``I bet one would
get the Virasoro formula''. 
This is because, with the particles integrated over the surface, there
is no cyclic order constraint, and any collection of particles are free
to approach each other and produce a singularity in $P^2$, where $P^\mu$
is the sum of their momenta. This is what happens in the Virasoro
formula\cite{Viras4pt}.

Should the fishnets, or electric fields, or currents, fill the ball,
or should they be confined to the surface with the external particles? 
In my view, the new Virasoro
identities were associated with the local conformal invariance of 
analytic functions, a much richer group than conformal transformations
in Euclidean spaces of higher dimension. 
Thus filling the three dimensional ball was unlikely to work, 
but putting fishnets on the surface might be very interesting.
So the three of us
began to work out electrostatics on the surface of the sphere.

In the Nielsen approach one needs the electrostatic energy of a 
configuration of point charges at arbitrary locations, and then integrate
over the charges' positions. 
We can solve Poisson's equation for each configuration of charges on a 
2-sphere, but we cannot define the electrostatic energy
as the integral of 
$(\vec \nabla \phi)^2$, because that includes the infinite
self-energy of each charge. Instead we might define 
$E=\half\sum_{i\neq j} q_i\phi_j(\vec r_i)$, but to do so one needs to
be able to 
find the electrostatic potential of a single charge.  We cannot
have a source of electric flux without a sink, and we seemed to hit an
impasse and let the matter drop. 
Several weeks later, after we had all gone our separate ways, while I 
was (I think) in Aspen, I
decided to look at this again, and I realized that putting an
arbitrary sink for all
the fields would do no harm. After all, for two-dimensional
electrostatics one takes $\phi \propto \ln |\vec r-\vec r_0|$ without
worrying about the flux which goes off to infinity, and in two
dimensions conformal invariance makes infinity no different from any
other point.
As the sphere is conformally equivalent to
the complex plane, the potentials are just logarithms of
$|(z_i,z_j,a,b)|$, the absolute value of the cross ratio, including
the arbitrary sink point $b$ and a point $a$ at which we can set the
potential to zero. The M\"obius invariance, now extended to three
{\em complex} parameters, again permits the positions of three charges to be
fixed arbitrarily, and the others integrated over the sphere, or
complex plane. We have 
consistency only for $\alpha(0)=2$, but there we have a consistent
n-point function for closed string\footnote{For some bizarre reason, I 
referred to the dual models in terms of what we would now call their 
world-sheets, as strip and tube models for what we would now call open
and closed strings.}
scattering\cite{ClosedStr}. I speculated that this
was equivalent to the Pomerons which appeared as a problem in the loop
graphs of open strings, and later, with Clavelli\cite{ClavShap}, I
showed that this is indeed the case.

There was at the time not much interest in closed strings, which have
no ends.
All the semi-successes of dual model phenomenology were based
on Harari-Rosner diagrams\cite{HRHarari,HRRosner} being incorporated
by Chan-Paton factors\cite{ChanPaton}, which required string ends 
on which to attach quarks. Even I postponed looking at factorization
and loop graphs in this model in favor of a paper\cite{JaSnonorient}
showing that 
nonorientable graphs do not enter theories with $SU(n)$ flavors incorporated
{\it \`a la} Chan-Paton. But the following summer, in Aspen again, 
with the factorization having been done by
others\cite{Yoshimura,DelGnDiV},
I addressed the one-loop diagram\cite{ClosedLoop}. 
The propagators now have an integral
over the length of the tube and the angle of twist to get to the next
particle, and the complex variable $w$ has $|w| = - \ln T$, where $T$
is the combined times of propagation, but $w$ also has a phase given by the
angle of twist in sewing the initial end of the tube to the final end.
The amplitude involves
$$ \int_{|w|\leq 1} d^2 w \prod_{r=1}^\infty |1-w^r|^{-2(D-E)},$$
where $D$ is the dimension of space-time and $E$ is the number of
factors assumed to mysteriously disappear if one removes the spurious
states. $E$ had already been shown to be 1 in general $D$, but we were
hoping, as was found true later, that $E=2$ in the right $D$. Still,
$D-E$ is a positive number, and $1-w^r$ vanishes for $w$ any integral
power of $e^{2\pi i/r}$,
so we have a terrible divergence at every point of the unit circle 
for which the angle is $2\pi$ times a rational number.

Fortunately, by the time I was looking at this, I had read the elegant
reformulation\cite{NeveuScherk} of the open-string loop in terms of 
Jacobi theta functions. This encourages us to look at $\tau = (\ln
w)/(2\pi i)$. Of course the integrand is invariant under
$\tau\rightarrow \tau+1$, because $w$ is unchanged, but it is also 
invariant under the Jacobi imaginary transformation, $\tau\rightarrow
-1/\tau$, provided we have the magic dimensions $D=26, E=2$. 
The world sheet of a loop of closed strings is a donut, conformally
equivalent to a parallelogram with sides 1 and $i\tau$, with opposite
sides identified. While a Hula hoop\texttrademark\ and a Mayflower
donut\cite{Donut} may not look the same, multiplying the parallelogram
by $-i/\tau$ maps one into the other.
These 
transformations generate the modular group, so invariance shows that in
integrating $w$ over the unit disk we are including an infinite number
of copies, while we wanted, for unitarity, only one copy of the region
around $w\approx 0$. Thus the right thing to do is restrict our
integration to the fundamental region, which is $|\tau|\geq 1$, 
$-\half < \re \tau \leq \half$. In terms of $w$, this is a subset of 
$|w|<0.0044$, so 
we stay far away from the horrible divergences.

\medskip\vskip -2pt\noindent
\begin{minipage}[h]{3.4in}\mpindent
If closed strings aroused little interest, loops of them really
aroused none. The figure shows the number of citations to \cite{ClosedLoop}
each year as listed by Spires. Interest in dual models as models
of the strong interactions was fading fast. Firstly, the evidence
for partons, pointlike constituents of hadrons, found in deep inelastic
scattering starting in 1969, was inconsistent with the soft, extended
object picture of strings. Secondly, 
non-abelian gauge theories were proven 
renormalizable ('t Hooft, 1971 \cite{tHooft}), 
explained neu-
\phantom{mmmmmmm}
\end{minipage}
\quad
\begin{minipage}[h]{2.8in}
\includegraphics[width=2.8truein]{loopcite.eps}\\
\phantom{mn} 
Citations by year to ref. \cite{ClosedLoop}, as \\
\phantom{mn} 
listed in the Spires Citation Index. 
\end{minipage}\\
\vskip-15pt\noindent
tral currents in a unified electroweak theory, and
gave quantum chromodynamics as a theory of the strong interactions.
This greatly improved the appeal of conventional field theory
at the expense of string theory. And within string theory, the inclusion
of fermions by Ramond\cite{Ramond} and Neveu-Schwarz\cite{NeveuSchwarz} was 
more exciting than loops of Pomerons.

In the fall on 1971 I started an Assistant Professorship at Rutgers
in a new high energy theory group headed by Lovelace, and including
Clavelli as a postdoc. Lovelace had a very ambitious program for 
describing arbitrary multiloop diagrams, and Clavelli and I looked at
how the closed string intermediate state in the nonplanar loop
interacts with the ordinary (open-string) states. Then I struggled
with understanding renormalization\cite{JaSrenorm}, an effort which
would have been 
totally trivial if I had only realized that $\sum_n n = -1/12$,
but unfortunately I had taken too many pure math classes to recognize
this fact.

The decision on my tenure was coming up, and dual models did not seem
the best way to prove my worth, so I reluctantly got into several
other endeavors, and was rather slow to get back into strings when 
they arose like a phoenix, or perhaps like a fire storm, in 1984. But
that is after the period we are considering here.

\section*{A Comment on Impact}

I want to say a few words about how this field was perceived within
the physics community. In recent years there have been numerous
attacks from some in the high energy theory community, and from
experimentalists, that strings are, like the Pied Piper, leading the
bright young theorists astray. String theory was not quite so dominant
in the 1969-1974 era, though it did absorb the attention of a very
large fraction of the young theorists. It did not get a similar
acceptance by most of the more established people, though I think Europe
was more receptive than America. In particular
Phys.~Rev.~Lett.~published very few articles in this field, but
Physics Letters had many of the important papers.
  
The field did attract quite a bit of attention. In fact, in the early '70's
I was interviewed by
a sociologist who wanted to do a study of what attracted so many
people to working on dual models. Unfortunately she followed a
narrow set of preprepared questions which seemed totally
off the mark. Her focus was on
which experimental data encouraged me to continue working
in the field. I don't think 
anything came of that study, and I haven't heard of studies done when
the field became even more rabid in the mid '80's.

But this is still an important question: Is there any real physics 
in string theory, and should so many people be working on
it. Undoubtedly there will be a shift towards more applied high energy
theory as LHC starts giving us more data to work with. It will be very
interesting to see where the field goes.

\end{document}